# Specific Characteristics of Applying the Paired Comparison Method for Parameterization of Consumer Wants

**Vasiliy Saiko**
**Institute for Entrepreneurship**
**"Strategy", Zhovty Vody, Ukraine**

**ABSTRACT.** The article describes the main problems concerned with using expert assessment method in consumer preference researches. The author proved the expediency of using a 3-point measurement scale. The author suggested an algorithm for controlling the judgments' consistency that includes analyzing and correcting the input estimates in real-time mode.

## 1 Problem definition

Forming the set of consumer wants (CW) and determining quantitative estimates of significance of elements in such a set constitute a key phase of several methods used in marketing and management. Quality Function Deployment («Quality house») [Aka90] can be an example. It uses quantitative interpretation of significance of CW in calculations for substantiation of choosing strategic decisions aimed at improving the product quality. Sources of information about CW are usually surveys as well as expert assessments.

However, detailed elaboration of CW for specific conditions and tasks often produces a large and unstructured list. In that case collecting estimates using usual surveys is substantially complicated, and expert assessment methods seem to be the most efficient.

One of the most popular expert assessment method was offered by Cohan, it is considered the most convenient way of getting reliable data in sociology and psychology. The expert's task is determining a more preferred object in every pair of objects presented to him one by one.

The quantity of pairs is n(n-1)/2, where n stands for the quantity of objects. The results are summarized in a table where frequencies of choosing objects i in each set as more preferred are calculated (C-frequencies, $c_i$). This method is classified as direct ranking, and considering limited potential of the used scale it provides extremely inaccurate idea on preference of a certain object. Indeed, supposing there is an object I for which $\neg \exists \{I, J \mid c_i > c_j\}$, then $c_i=0$. Thus, quantitative interpretation of preference in this case is not complete.

A more informative scale – focusing scale – can be used with help of ranking procedure according to Louis Thurstone [Thu27]. Object I when preferred over J receives the estimate 1 ($a_{ij}=1$), and when not, it receives the estimate 0. If the expert considers the objects equal they both receive equal estimates 0.5 ($a_{ij}=0.5$). In the matrix where the comparison results are summarized the main diagonal stays void.

If after the assessment carried out by k experts they come up with empirical frequency $f_{ij}$ corresponding to the number of preferences of I over J, then the intensity of such preference $p_{ij}$ is supposed to be equal to $f_{ij}/k$. In addition, Thurstone postulates that $p_{ij}$ is normally distributed.

The above mentioned matrix of paired comparison must meet the asymmetry condition (if $a_{ij}=1$ then $a_{ji}=0$) and the transitivity condition (if $a_i > a_m$ and $a_m > a_j$ then $a_i > a_j$). However, it is impossible to ensure strict following of these rules, and in sociological research practice there is an agreement: if the initial matrix has few violations of asymmetry and transitivity then it is possible to apply the method. It is possible to determine the critical barrier more accurately only based on practical experience of a researcher [Tol98].

The methods considered above have a substantial drawback – they do not provide accurate estimates of significance (importance) of the compared objects because they do not ensure the transitivity of the recorded judgments. So, they are used exclusively for tasks of ranking objects or parameters.

Further development of paired assessment method is connected with T. Saaty. In 1978 he offered the Analytic Hierarchy Process (AHP) [Saa78] which subsequently became a kind of standard. AHP has been applied successfully in many different spheres: military science, climatology, economics etc.

The main component of AHP is a matrix of judgments $\mathbf{A} = (a_{ij})$, i,j=1,2,…,h, where $a_{ij}$ is a number corresponding to significance of the object I compared to J.

According to the scale of T. Saaty, $a_{ij}$ can possess values from 1 (I has the same meaning as J) to 9 (I is much more important than J) or to 1/9 (I is much less important than J). This data is entered to the matrix above the main diagonal. The main diagonal is filled with numbers one. Then a reversely symmetrical matrix is made, and the required weight vector is determined as an eigenvector of this matrix correspondent to the maximum eigenvalue $l_{max}$. Indeed, the weight vector is an eigenvector of a consistent matrix correspondent to its maximum eigenvalue h.

However, when making a paired comparison matrix according to AHP it is almost impossible to ensure its consistency. For controlling accuracy of the results T. Saaty introduced a special index called consistency index. They also determine stochastic consistency coefficient $RI = 1,98(h-2)/h$ and consistency coefficient $CR = CI/RI$. Coefficient CR must not exceed 0,1. Otherwise the expert's estimates are advised to be revised.

Thus, applying the paired comparison method for collecting and analyzing information about CW is connected with two main problems described below.

## 1.1 Choosing and substantiation of the measure scale that would minimize errors caused by indistinctness of the information representation

Judgments of the experts are not numerical. However, they have to express them in numbers. Therefore, choosing a comprehensive and informative scale takes critical importance, especially when not so much ranking of the assessment objects as calculating their significance is required.

Models of the experts' behavior are based on the assumption that the parameters are assessed with certain errors. The most often mentioned reasons for errors are: incompleteness of knowledge about the objects' characteristics, insufficient confidence of the expert in his judgments' accuracy, contradictory knowledge, indistinctness of the information representation.

When comparing large quantities of parameters having different nature, using 9-point assessment scale inevitably causes complications. Meanwhile, in research practice assessing the compared objects in the following terms "the objects are equal"-"object 1 is more important than object 2"-"object 1 is much more important than object 2" seems to be much more relevant. In the method of Saaty such a scale can be represented by a set $\{1/G, 1/F, 1, F, G\}$, where F and G are integer, and $G>F$.

For contrastive analysis of the accuracy of the suggested scale and scale of T. Saaty experiments have been carried out involving assessing «weights» of a random set of 10 positive integers from 1 to 10 with different values of F and G. In the main, the experiments were carried out under circumstances excluding the first four sources of errors, coefficients $k_i = i \Big/ \sum_{i=1}^{10} i$ were taken as true estimates of the objects.

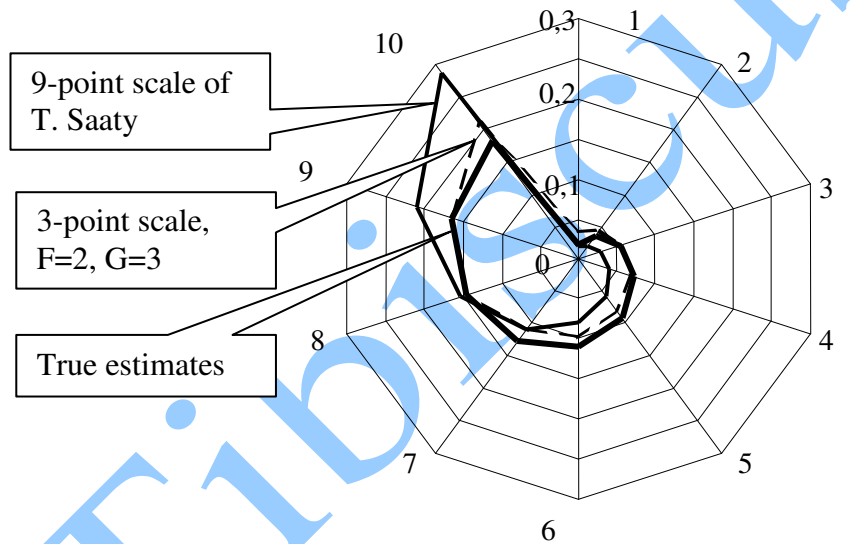

Fig. 1. Contrastive analysis of accuracy of paired comparison methods using 9-point scale of T. Saaty and 3-point scale.

The comparison results represented on fig.1 show that measurement error when using the 3-point scale is significantly lower than in case of using the 9-point scale of Saaty.

## 1.2 Excluding measurement faults caused by intransitivity of judgments and automatic errors

At the moment the task of automating the control of consistency and transitivity of the experts' judgments is urgent. The work [Bel07], for instance, offers a method of discrete minimization of inconsistency of the expert data based on determining their numeric value. Search for the most erroneous estimates is carried out in the data fragments substantially affecting the whole array of estimates and determining the integral estimate value – consistency relation CR. The criterion for decision making and choosing the best alternative is minimizing the numeric value of the error index. The algorithm for the error correction suggested in [Har99] works with a complete paired comparison matrix and allows us to determine estimates characterized by maximum value of the inconsistency relation more accurately with the help of estimates characterized by minimum value of the inconsistency relation. However, the suggested method of correcting errors without experts' participation does not seem to be universal and proper. In particular, the author does not mention the problem of an "off-scale reading" of transitive closure, i.e. when, for instance, $a_{im}>5$ and $a_{mj}>5$, then the estimate $a_{ij}$ must exceed the scale range.

The algorithm of providing coherence of the estimates described in [Bel07] is also connected with the analysis of prepared data. It includes building an auxiliary matrix where data of preliminary judgments' examination is automatically entered. In case there are differences between the data in auxiliary and main matrices the data is either deleted or is produced to the expert once more.

The common feature of the consistency control methods under consideration is using a ready judgment matrix and automatic correction of estimates that does not always ensure preservation of significant data. Besides, it does not provide eliminating "automatic" input errors often causing inaccurate results.

In the course of experiments carried out by the author of this research with matrices of large dimensions (h = 20…30) he found out that substituting one of the judgments with its exact opposite can cause, for example, a 7% change of CI but this also causes a 30% value change of one of the final weighting coefficients. Thus, the paired comparison method is extremely sensitive to errors, and no control method can ultimately solve this problem yet.

## 2 Results

The purpose of the present research is developing algorithms and software for collecting and processing experts' estimates within the paired comparison method. It is assumed that the list of the objects being researched is large enough (exceeding 20 items) and is produced to qualified consumers who are not "experts" in popular sense. As the most promising method for controlling the judgments' consistency the author suggests organizing transitivity analysis in real-time mode. In case of revealing an error an expert must be offered to correct the erroneous judgment shown to him.

Let us consider a judgment matrix concerning h objects $A_k = (a_{ij})_{h \times h}$, h>20. In the course of work the k-th expert assesses one by one the parameter 1 and 2,….,j, then parameter 2 and 3,…,j, and so on. The last pair he compares includes parameters (i-1)-th and j-th. The received estimates $a_{ij}$ are entered above the main diagonal of the matrix. The matrix below the main diagonal is filled with estimates that equal $1/a_{ij}$.

The paired comparison is executed in terms of domination of one element over another measured with a 3-point scale (see fig. 2). Calculation of final weighting coefficients' values is executed as follows: one calculates the sums in columns and divides the elements of each column by the corresponding sum. The elements of the final column matrix of weighting coefficients are calculated as mean values of the coefficients determined during the previous phase for each row.

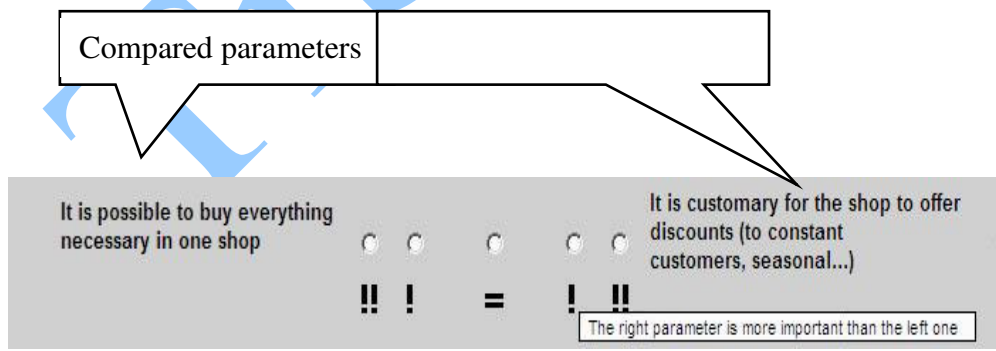

Fig. 2. Interface fragment of the paired comparison software.

Data concerning the experts' estimates is accumulated, and as the accumulation goes on mean values of the weighting coefficients in question are determined more accurately. The quantity of experts is determined by the required size of confidence interval.

The suggested transitivity control algorithm is based on the assertion that for each estimate $a_{ij}$ expressed by an expert starting from i=2 the elements $a_{m,j},…, a_{1,j}$ connected with the current estimate of the i-th and j-th parameters, such that m=i-1,..,1, and elements $a_{i,m},…, a_{i,1}$ are known. Thus, starting from the second row of the matrix there is a possibility to control transitivity of the triads shown on fig. 3. At the same time 27 variants for combinations of preference/equality relations in each triad under consideration are possible.

Systematizing the variants of intransitivity rise allowed to reveal 14 possible conflict situations (table 1) and to proceed to algorithmization of the procedure of "conflict" triads' identification.

It is noteworthy that only while analyzing the 2nd row one should search for an error in an estimate of any of the pairs I-J, M-J and I-M. After checking the elements of the first two rows, starting from the 3rd row an error can be contained only in the estimate of the current pair I-J, other pairs are considered checked and corrected, if necessary.

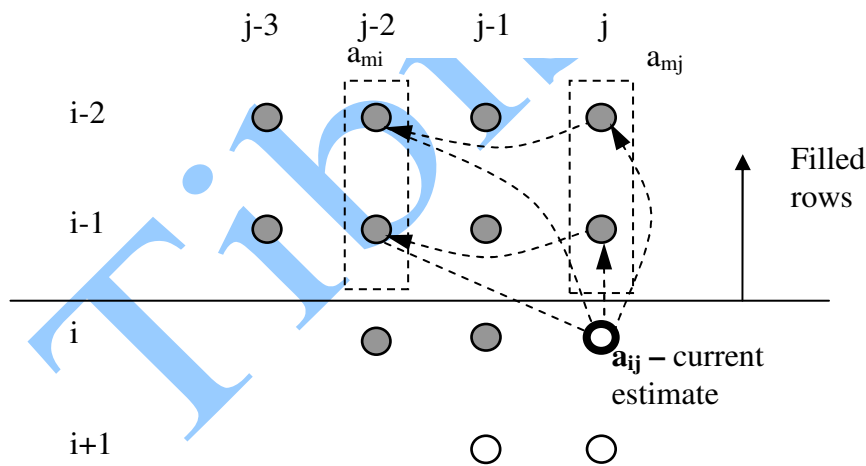

Fig. 3. Choosing elements of matrix for transitivity analysis.

Table 1.
Identifying intransitive judgment triads

| Assessment result M -J | Current assessment result I – J | | | | | |
|---|---|---|---|---|---|---|
| | $a_{ij} > 1$ | | $a_{ij} < 1$ | | $a_{ij} = 1$ | |
| $a_{mj} > 1$ | - | - | $a_{mi} > 1$ | $a_{mi} = 1$ | $a_{mi} < 1$ | $a_{mi} = 1$ |
| $a_{mj} < 1$ | $a_{mi} > 1$ | $a_{mi} = 1$ | - | - | $a_{mi} > 1$ | $a_{mi} = 1$ |
| $a_{mj} = 1$ | $a_{mi} > 1$ | $a_{mi} = 1$ | $a_{mi} < 1;$ | $a_{mi} = 1$ | $a_{mi} < 1$ | $a_{mi} > 1$ |

The algorithm's efficiency when using large matrices is illustrated by diagrams shown on fig. 4 and 5. In both cases calculating of weighting coefficients of the assessed characteristics was carried out by the same three experts. Size of the confidence interval (thin lines) in the second case shows that these estimates are more consistent (CR=0,02-0,05).

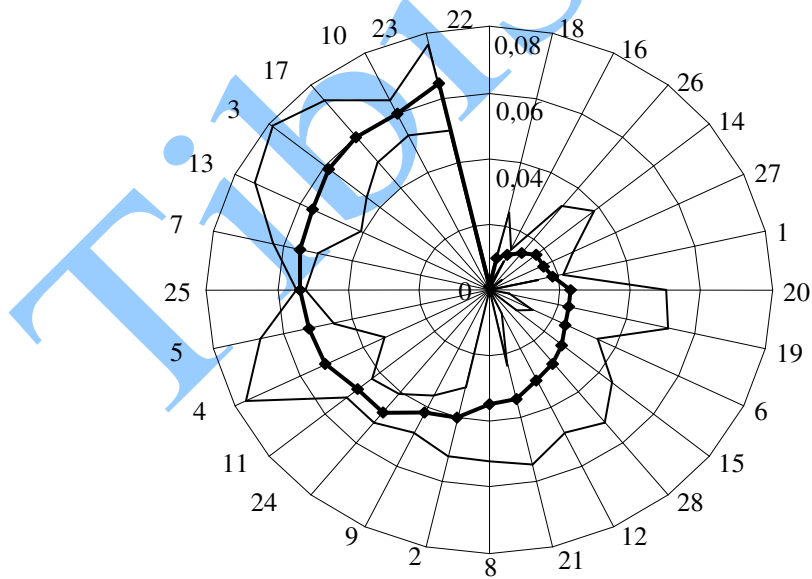

Fig. 4. Examination result without transitivity control
(3 experts, 28 assessed parameters).

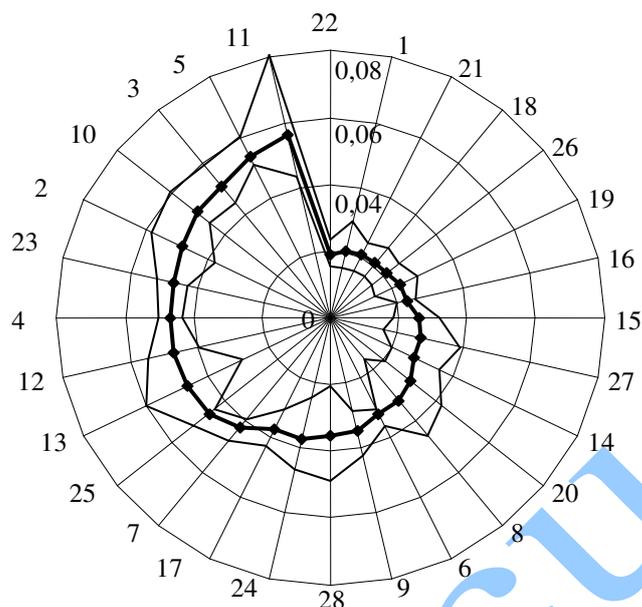

Fig.5. Examination result with transitivity control
(3 experts, 28 assessed parameters).

## Conclusions

The completed research revealed main problems concerned with using expert assessment method in large consumer preference researches. The author proved the expediency of using a 3-point scale. The experiments showed that the scale "the objects are equal" - "object 1 is more important than object 2" - "object 1 is much more important than object 2" ensures sufficient accuracy under circumstances of comparing large quantity of objects having different nature. The author suggested an algorithm for controlling the judgments' consistency that includes analyzing and correcting the input estimates in real-time mode. The developed software (VBA, Excel) is currently used in teaching process.

**References**

[Aka90]   Yoji Akao (Ed.) *Quality Function Deployment (QFD). Integrating Customer Requirements into Product Design* - Portland, OR: Productivity Press, 1990. - 369 p.

[Bel07]   К. Н. Белоусов - Алгоритмическое обеспечение проверки непротиворечивости экспертных оценок. Фундаментальные исследования. -2007.-№12. с.37-38.

[Har99]   Е. В. Харитонов - Согласование исходной субъективной информации в методах анализа иерархий. Математическая морфология. - Т. 3. - Вып. 2 – 1999. – С. 41-51.

[Saa78]   T. L. Saaty - *Eploring the interface between hierarchies, multiple objectives and fuzzy sets. Fuzzy sets and Systems*, 1978, V.1. P.57-68.

[Thu27]   L. L. Thurstone - *A Law of Comparative Judgment, Psychological Review*, vol. 34.- 1927. pp. 273-286.

[Tol98]   Ю. Н. Толстова - Измерение в социологии. М.: Инфра-М.- 1998.- 178с.